\documentclass[onecolumn,useAMS,usenatbib,floats]{mn2e}
\usepackage{graphicx}
\usepackage{amsmath}
\usepackage{amssymb}
\usepackage{lscape}
\usepackage{color}
\usepackage{bm}
\usepackage{dsfont}
\usepackage{afterpage}
\usepackage{float}

\def\be{\begin{equation}}
\def\ee{\end{equation}}
\def\bea{\begin{eqnarray}}
\def\eea{\end{eqnarray}}

\def\wde{w_{de}}
\def \ncal{\mathcal{Q}}

\renewcommand{\a}{{A}}
\newcommand{\h}{\mathcal{H}}
\newcommand{\comment}[1]{}

\title{Adiabatic initial conditions for perturbations in interacting dark energy models}

\author[Elisabetta Majerotto, Jussi V\"{a}liviita and Roy Maartens]{Elisabetta Majerotto$^{1,2}$, Jussi V\"{a}liviita$^{2}$ and Roy Maartens$^{2}$\\
$^{1}$INAF-Osservatorio Astronomico di Brera, Via Bianchi 46, I-23807 Merate (LC), Italy\\
$^{2}$Institute of Cosmology \& Gravitation, University of Portsmouth, Portsmouth PO1 3FX, United Kingdom}

\begin{document}
\label{firstpage}

\date{Accepted 2009 December 1. Received 2009 November 13; in original form 2009 September 19}
%\date{Received in original form 2009 September 19}
\pagerange{\pageref{firstpage}--\pageref{lastpage}} \pubyear{20XX}

\maketitle

\begin{abstract}

We present a new systematic analysis of the early radiation era
solution in an interacting dark energy model to find the adiabatic
initial conditions for the Boltzmann integration. 
In a model where the interaction is proportional to the dark matter density,
adiabatic initial conditions and viable cosmologies are possible if the
early-time dark energy equation of state parameter is $w_e > -4/5$.
We find that when adiabaticity between cold dark matter,
baryons, neutrinos and photons is demanded, the dark energy
component satisfies automatically the adiabaticity condition. As
supernovae Ia or baryon acoustic oscillation data require the
recent-time equation of state parameter to be more negative, we
consider a time-varying equation of state in our model. In a
companion paper [arXiv:0907.4987] we apply the initial conditions
derived here, and perform a full  Monte Carlo Markov Chain
likelihood analysis of this model.

\end{abstract}

%\pacs{98.70.Vc, 98.80.Cq}

\maketitle

\section{Interacting dark energy}

Dark energy and dark matter, the dominant sources in the
`standard' model for the evolution of the universe, are currently
only detected via their gravitational effects. This implies an
inevitable degeneracy between them.  A dark sector interaction
could thus be consistent with current observational constraints.
We look at such a model, assuming that the dark matter and dark energy can be
treated as fluids whose interaction is proportional to the dark matter density.

For interacting dark energy, the energy balance equations in the
background are \bea
\label{eq.cont_rhoc} \rho_c' &=& -3 \h \rho_c + a Q_c\,, \\
\rho_{de}' &=& -3 \h (1+\wde) \rho_{de} + a Q_{de}\,,\quad\quad
Q_{de}=-Q_c\,, \label{eq.cont_rhode} \eea where  $\h = a'/a$,
$\wde=p_{de}/\rho_{de}$ is the dark energy equation of state
parameter, a prime indicates derivative with respect to conformal
time $\tau$, and $Q_c$ is the rate of transfer of dark matter
density due to the interaction.

Various forms for $Q_c$ have been investigated (see,
e.g.~\citet{Wetterich:1994bg, Amendola:1999qq, Billyard:2000bh, 
Zimdahl:2001ar, Farrar:2003uw, Chimento:2003iea, Olivares:2005tb, Koivisto:2005nr, Guo:2007zk, Sadjadi:2006qp, Boehmer:2008av, He:2008tn, Quartin:2008px, Pereira:2008at, 
Quercellini:2008vh, Valiviita:2008iv, He:2008si, Bean:2008ac, Chongchitnan:2008ry, 
Corasaniti:2008kx, CalderaCabral:2008bx, Gavela:2009cy, Jackson:2009mz}).
\emph{We consider} models where the interaction has the form of a decay of one species
into another - as in simple models of reheating and of curvaton decay
\citep{Malik:2002jb, Sasaki:2006kq, Assadullahi:2007uw}.
Such a model was introduced by
\citet{Boehmer:2008av, Valiviita:2008iv}. It is not derived from a Lagrangian
[in contrast with e.g. \citet{Wetterich:1994bg, Amendola:1999qq}], but it is motivated
physically as a simple phenomenological model for decay of dark matter
particles into dark energy. In this sense, it improves on most other
phenomenological models, which are typically designed for mathematical
simplicity, rather than as models of interaction. The methods that we
use here and in the companion paper \citep{CompanionLike} may readily
be extended to other interactions, including those based on a Lagrangian.
We assume that in the background the interation takes the form
\citep{Boehmer:2008av, Valiviita:2008iv} 
\be Q_c =- \Gamma
\rho_c \,,\label{eq:ourcoupling} 
\ee 
where $\Gamma$ is the
constant rate of transfer of dark matter density. Positive
$\Gamma$ corresponds to the decay of dark matter into dark energy,
while negative $\Gamma$ indicates a transfer of energy from dark
energy to dark matter.

In \citet{Valiviita:2008iv} we considered the case of fluid dark
energy with a constant equation of state parameter $-1 < \wde \le
-4/5$, and found a serious large-scale non-adiabatic instability
in the early radiation era. This instability grows stronger as
$\wde$ approaches $-1$. Phantom models, $\wde < -1$, do not suffer
from this instability, but we consider them to be unphysical.

The instability is determined by the early-time value of $\wde$.
We will show that the models are not affected by the large-scale
non-adiabatic instability during early radiation domination if at
early times $w_{de}> -4/5$. If we allow $\wde$ to vary, such a
large early $\wde$ can be still consistent with Supernovae Type Ia (SN) and baryon acoustic oscillation (BAO)
observations, provided that at late times $\wde \sim -1$.
In this paper, we represent $\wde$ via the
parametrization $ \wde = w_0 + w_a (1-a)$ \citep{Chevallier:2000qy, 
Linder:2002et}, which we rewrite as 
\be 
\wde = w_0a + w_e (1-a)\,,
\label{eq:wdeedef} 
\ee 
where $w_e =w_0+w_a$ is the early-time
value of $\wde$, while $w_0$ is the late-time value.

There are two critical features of the analysis of interacting
models, which are not always properly accounted for in the
literature:
\begin{itemize}
\item
The background energy transfer rate $Q_c$ does not in itself
determine the interaction in the perturbed universe. One should
also specify the momentum transfer rate, preferably via a physical
assumption. We make the physical assumption that the momentum
transfer vanishes in the dark matter rest-frame; this requires
that the energy-momentum transfer rate is given covariantly by
\citep{Kodama:1985bj, Valiviita:2008iv}
\be \label{qcmu}
Q_c^\mu = Q_c u_c^\mu = -Q_{de}^\mu\,,\quad\quad Q_c=-\Gamma
\rho_c(1+ \delta_c),
\ee
where $u_c^\mu$ is the dark matter 4-velocity, 
and $\delta_c=\delta\rho_c/\rho_c$ is the
cold dark matter (CDM) density contrast.
\item
Adiabatic initial conditions in the presence of a dark sector
interaction require a very careful analysis of the early-radiation
solution, both in the background and in the perturbations. We
derive these initial conditions by generalizing the methods of
\citet{Doran:2003xq} to the interacting case, thereby extending our
previous results \citep{Valiviita:2008iv}.
\end{itemize}

We give here the first systematic analysis of the initial
conditions for perturbations in the interacting model given by
Eq.~(\ref{qcmu}) -- and our methods can be adjusted to deal with
other forms of interaction. In the companion paper
\citet{CompanionLike} we report the results of our full  Monte Carlo Markov
Chain likelihood scans for this model. 
Cosmological perturbations of other interacting models have been
investigated, e.g., in
\citet{Amendola:2002bs, Koivisto:2005nr, Olivares:2006jr, Mainini:2007ft, 
Bean:2007ny, Vergani:2008jv, Pettorino:2008ez,  LaVacca:2008kq,
Schaefer:2008qs, Schaefer:2008ku, He:2008si, Bean:2008ac, 
Corasaniti:2008kx, Chongchitnan:2008ry, Jackson:2009mz, Gavela:2009cy, 
LaVacca:2009yp, He:2009mz, CalderaCabral:2009ja, He:2009pd, 
Koyama:2009gd, Kristiansen:2009yx}.

\section{Perturbation equations}

The scalar perturbations of the spatially flat Friedmann-Robertson-Walker metric are given by 
\be
\label{eq.gen gauge} 
ds^2 = a^2\Big\{ -(1+2\phi)
d\tau^2+2\partial_iB\,d\tau dx^i + \Big[(1-2\psi)\delta_{ij}+
2\partial_i\partial_j E\Big]dx^idx^j\Big\}\,. 
\ee
 In the perturbed
universe, the dark sector interaction involves a transfer of
momentum as well as energy. The covariant form of energy-momentum
transfer for a fluid component $A$ is $\nabla_\nu T^{\mu\nu}_{A } = Q^\mu_{A }$, where
$Q^\mu_{c}= a^{-1}(Q_c,\vec 0\,) =-Q^\mu_x$ in the background. The
perturbed energy-momentum transfer 4-vector can be split as
\citep{Valiviita:2008iv}
\be
Q^{A }_0 =-a\left[ Q_A(1+\phi) + \delta Q_A \right], \quad \quad
Q^{A }_i = a\partial_i\left( f_A- Q_A \frac{\theta}{ k^2} \right),
\label{Qmomentum}
\ee
where $k$ is the comoving wavenumber, $f_\a$ is the intrinsic momentum transfer potential and
$\theta= (\rho+p)^{-1}\sum (\rho_A+p_A)\theta_A$ is the total
velocity perturbation ($ \theta=-k^2v$). The evolution equations
for density perturbations and velocity perturbations for a generic
fluid are \citep{Valiviita:2008iv, Kodama:1985bj}
 \bea
&&\delta_\a'+3{\cal H}(c_{s\a}^2-w_\a)\delta_\a
+(1+w_\a)\theta_\a+ 3{\cal H}\big[3{\cal
H}(1+w_\a)(c_{s\a}^2-w_\a)+w_\a' \big]
\frac{\theta_\a }{ k^2} \nonumber\\
&&~~-3(1+w_\a)\psi'+ (1+w_\a)k^2\big(B-E'\big) =\frac{aQ_\a }{
\rho_\a}\left[ \phi-\delta_\a+3{\cal H}(c_{s\a}^2-w_\a)
\frac{\theta_\a }{ k^2} \right] +\frac{a}{ \rho_\a}\,
\delta Q_\a\,,\label{dpa}\\
&& \theta_\a'+{\cal H}\big(1-3c_{s\a}^2\big)\theta_\a-
\frac{c_{s\a}^2 }{ (1+w_\a)}\,k^2\delta_\a +\frac{2 w_A}{ 3
(1+w_\a)}\,k^2 \pi_\a -k^2\phi = \frac{aQ_\a }{ (1+w_\a)\rho_\a}\big[ \theta-
(1+c_{s\a}^2)\theta_\a \big] -\frac{a}{ (1+w_\a)\rho_\a} \,
k^2f_\a\,. \label{vpa}
 \eea
%\nonumber\\
%&&~~~
Here $c_{s\a}^2$ is the sound
speed, and $\pi_\a$ is the anisotropic stress. For our model
$\pi_{de}=0$, and we set $c_{s\, de}^2 = 1$, as in standard
non-interacting quintessence models, in order to avoid adiabatic
instabilities (see discussion in \citet{Valiviita:2008iv}).

For the interaction defined by Eq.~(\ref{qcmu}), we find from
Eq.~(\ref{Qmomentum}) that
 \be
f_c= \Gamma \frac{\rho_c}{k^2} \left( \theta_c - \theta \right) =
-f_{de}\,. \ee Then we can write the dark energy and cold dark
matter perturbation equations for our model:
 \bea
\label{eq.delta'de_ourQ} && \delta'_{de} + 3\mathcal
H(1-w_{de})\delta_{de} + (1+w_{de})\left[\theta_{de} +k^2
(B-E')\right] + 9\mathcal H^2(1-w_{de}^2)\frac{\theta_{de}}{k^2} -
3 a \h^2 w_a \frac{\theta_{de}}{k^2}
-3(1+w_{de})\psi' = \nonumber\\
&& \quad \quad a\Gamma \frac{\rho_c}{\rho_{de}}\left[\delta_c
-\delta_{de} + 3\mathcal
H (1-w_{de})\frac{\theta_{de}}{k^2}+\phi\right]\!, \\
&& \theta'_{de} -2 \mathcal H\theta_{de}
-\frac{k^2}{(1+w_{de})}\delta_{de} - k^2\phi =
\frac{a\Gamma}{(1+w_{de})}\frac{\rho_c}{\rho_{de}}
\left(\theta_c-2\theta_{de}
\right)\,, \label{eq.theta'de_ourQ}\\
\label{eq.delta'c_ourQ} && \delta'_c +\theta_c + k^2 (B-E')
-3\psi' =-
a\Gamma\phi\,, \\
&& \theta'_c + \mathcal H \theta_c -k^2\phi = 0\,.
\label{eq.theta'c_ourQ}
 \eea

\section{Background solution in early radiation era}

The background solution in the early radiation era
($\rho_{\mathrm{tot}} \simeq \rho_r$) is important for finding the
initial conditions for the integration of cosmological
perturbations. In what follows we use occasionally the Hubble
parameter $H=a^{-1}\h$ instead of the conformal Hubble parameter
$\h$. In the radiation era we have \be \h = \tau^{-1}\ \ \ \
\mbox{ and } \ \ \ \ a = \h_0\sqrt{\Omega_{r0}}\,\tau\,,
\label{eq:confHatRD} \ee where $\h_0$ is the conformal Hubble
parameter today, and $\Omega_{r0} =
\rho_{r0}/\rho_{\mathrm{crit}0} \approx 2.47\times 10^{-5} h^{-2}$
is the radiation energy density parameter today. Here $h$ is
defined by $H_0 = h\times100\,$km$\,$s$^{-1}$Mpc$^{-1}$, and as
$a_0=1$, we have $H_0 = \h_0$. Furthermore, we have $H =
(2t)^{-1}$ and $a = (2H_0\sqrt{\Omega_{r0}}\,t)^{1/2}$ where $t$
is the cosmic time. By Eq.~(\ref{eq:confHatRD}) we find \be t =
(H_0 \sqrt{\Omega_{r0}}\,/2)\tau^2\,. \label{eq:tvstau} \ee

We define the ratio of dark energy to cold dark matter density $ r
= {\rho_{de}}/{\rho_c}$. Then, employing
Eqs.~(\ref{eq.cont_rhoc}) and (\ref{eq.cont_rhode}),
\be \dot{r} =
-\left\{ \frac{3 w_e}{2t} - \left[ \Gamma +3 w_{a} \left( \frac{H_0
\sqrt{\Omega_{r0}}}{2t} \right)^{1/2}\right] \right\} r +
\Gamma\,,\label{eq.rdot}
\ee
where the dot indicates derivative with respect to cosmic time. At early enough times, $|\Gamma + 3
w_{a} (H_0 \sqrt{\Omega_{r0}}/2t)^{1/2}| \ll 3 | w_e |/(2 t)$, and
we can neglect the term in square brackets, so that the
solution is \be r = r_{\mathrm{ref}}
\left(\frac{t}{t_{\mathrm{ref}}}\right)^{-3 w_e/2} + \frac{2
\Gamma}{3 w_e + 2}t\,, \label{eq.early_bg_fullsol} \ee where
$r_{\mathrm{ref}}$ is an integration constant corresponding to
$\rho_{de}/\rho_c$ at some (early) reference time
$t_{\mathrm{ref}}$ in the case where $\Gamma=0$. From Eq.
(\ref{eq.early_bg_fullsol}) we find that we have two regimes,
depending on the value of the early-time equation of state
parameter $w_e$. If $w_e \le-2/3$, then the second term dominates
over the first as $t\to 0$, and we recover the solution of
\citet{Valiviita:2008iv}: \be \frac{\rho_{de}}{\rho_c} = \frac{a
\Gamma}{3 w_e +2}\h^{-1} = \tilde C (k\tau)^2\,, \quad\quad \tilde
C = \frac{H_0\Gamma}{k^2} \frac{\sqrt{\Omega_{r0}}}{3 w_e +2}\,.
\label{eq:blowuprevol} \ee
If $w_e>-2/3$, then the first term in
Eq.~(\ref{eq.early_bg_fullsol}) dominates: \be
\frac{\rho_{de}}{\rho_c} \simeq r_{\mathrm{ref}}
\left(\frac{t}{t_{\rm ref}}\right)^{-3 w_e/2} = C (k\tau)^{-3
w_e}\,, \quad\quad C = r_{\mathrm{ref}} \left( \frac{H_0
\sqrt{\Omega_{r0}}}{2t_{\mathrm{ref}}k^2}\right)^{-3 w_e/2}\,.
\label{eq.r_evol} \ee

For the background evolution of $\rho_c$ in the radiation
dominated era, \be \dot{\rho}_c = - \left( \frac{3}{2 t} +  \Gamma
\right) \rho_c\,, \label{eq:timerhoc} \ee the second term in
brackets is negligible relative to the first at times
$t\ll3/(2\Gamma)$, or $\tau \ll \tau_{\mathrm{switch}} =
({\sqrt{\Omega_{r0}}} {|\Gamma/H_0|}/3)^{-1/2}H_0^{-1}$. For these
times, $\rho_c \propto a^{-3}$. In typical models that provide a
good fit to Cosmic Microwave Background (CMB) data, $H_0^{-1} = {\cal O}(10)\,$Gpc, and the
conformal time at matter-radiation equality is $\tau_{eq} = {\cal
O}(100)\,$Mpc. If we demand that the evolution of $\rho_c$ is
effectively standard during the whole radiation dominated era,
i.e. $\tau_{\mathrm{switch}}>\tau_{eq}$, we require \be
\left|\frac{\Gamma}{H_0}\right| \lesssim
\frac{30\,000}{\sqrt{\Omega_{r0}}} \approx 10^6 \, h\,, \ee where
$h\sim 0.7$. As we study in this paper coupling strengths
$|\Gamma/H_0| \lesssim 1$, we can safely assume that the cold dark matter
evolution during radiation domination is completely the standard
non-interacting one $ \rho_c = \rho_c^{eq} ({a}/{a_{eq}})^{-3}$,
where $\rho_c^{eq}$ and $a_{eq}$ are the dark matter energy density and
the scale factor at matter-radiation equality, respectively.
Noticing that the radiation energy density can be written as
$\rho_r = \rho_r^{eq}(a/a_{eq})^{-4}$, and that $\rho_c^{eq} =
\rho_r^{eq}$ by definition, we find that in the radiation
dominated era \be \frac{\rho_c}{\rho_r} = \frac{a}{a_{eq}} =
\omega_2 \,k\tau\,,\quad\quad \omega_2 =
\frac{H_0}{k}\frac{\sqrt{\Omega_{r0}}}{a_{eq}},
\label{eq:rhocorhor} \ee where we used Eq.~(\ref{eq:confHatRD}). In
the non-interacting case we could continue by setting $a_{eq}
= \Omega_{r0}/\Omega_{c0}$, but in the interacting case the dark matter
evolution from $\tau_{\rm switch}$ up to today ($\tau_0$) differs
from $\propto a^{-3}$: by Eq.~(\ref{eq:timerhoc}), it follows that
at recent times, for a positive $\Gamma$, the dark matter density decreases
faster, and with a negative $\Gamma$ it decreases slower than
$a^{-3}$. Therefore we cannot do the ``$a^{-3}$ scaling'' all the
way up to today, but instead have to stop at some early enough
reference time. Here we choose the time of matter-radiation
equality.

An upper limit to the early dark energy (DE) equation of state $w_e$ could be
set by requiring dark matter domination over DE at early times. Then
Eq.~(\ref{eq.r_evol}) would set the constraint $w_e<0$. However,
if the DE equation of state is close to $0$ at early times, it
could well mimic the behaviour of cold dark matter. On the other hand, if $w_e$
is close to $1/3$, the ``DE'' component would behave like radiation
at early times. So, for $0 \leq w_e < 1/3$, we conclude that the
fluid which at late times behaves like dark energy, behaves at
early times like a combination of matter and radiation. As this
case cannot be ruled out, we set a conservative upper bound on
$w_e$ by demanding that in the early universe DE does not dominate
over radiation, i.e., for $\tau \to 0$,  we have $\rho_{de}/\rho_r
\to 0$. Using Eqs.~(\ref{eq.r_evol}) and (\ref{eq:rhocorhor}), we
find 
\be \label{eq.omega_de_evol} 
\frac{\rho_{de}}{\rho_r} =
\frac{\rho_{de}}{\rho_c} \, \frac{\rho_c}{\rho_r} = C \omega_2
(k\tau)^{1-3w_e}\,,
\ee 
which implies, as expected, $w_e < 1/3$.

\section{Super-Horizon initial conditions for perturbations}

In order to solve numerically the perturbation equations we need
to specify initial conditions in the early radiation era. The
wavelength of the relevant fluctuations is far outside the horizon
during this period: $k\tau \ll 1$. To compute the initial
conditions we start by writing the perturbation equations of each
species and the perturbed Einstein equations in terms of the
gauge-invariant variables developed by
\citet{Bardeen:1980kt}:
\begin{eqnarray}
\label{eq.PhiandPsi} && \Phi = -\psi + \h(B-E') \,, \quad \quad
\quad
\Psi =  \phi + \h \left( B-E' \right) + \left( B-E'\right)' \nonumber\,,\\
\label{eq.GI_denandvel} && \Delta_A = \delta_A +
\h^{-1}\frac{\rho_A'}{\rho_A} \psi\,,  \quad \quad \quad V_A =
k^{-1}\theta_A + k(B -E')\,, \quad \quad \quad \Pi_A = \pi_A .
\label{eq.conversion}
\end{eqnarray}
The general evolution equations for the density, velocity and anisotropic stress perturbations $\Delta_A$,  $V_A$ and $\Pi_A$ and the
Einstein equations for the metric perturbations $\Phi$, $\Psi$ are given
in~\citet{Kodama:1985bj}.

We use and generalize the systematic method of \citet{Doran:2003xq}
in order to analyze the initial conditions in the interacting DE
model with time-varying $w_e$. The results are derived below, but
let us summarize the key points before going to the details. The
conclusion is that we can use adiabatic initial conditions for
 \be
-\frac{4 }{ 5} \le w_e \le -\frac{2 }{ 3} \quad\quad \mbox {or}
\quad\quad -\frac{2 }{ 3} < w_e < \frac{1}{ 3} \,.
 \ee
For both of these intervals, the initial conditions for all
non-dark energy quantities are the same as in the non-interacting
case. For the second $w_e$ interval, the initial dark energy
density perturbation is the same as the standard one,
$\Delta_{de}=3(1+w_e)\Delta_\gamma/4$, whereas for the first
interval, we find a non-standard initial condition $
\Delta_{de}=\Delta_\gamma/4$. The difference arises because of the
different background evolution in the two cases (as given in the
previous section). Note that for $w_e < -1$ it is also possible to
have adiabatic initial conditions, but we consider this case to be
unphysical. For $-1 \le w_e \le -4/5$, we recover the
non-adiabatic blow-up case of \citet{Valiviita:2008iv}.

Similar considerations could be extended to the early matter
dominated era. The key difference there is that the background
behaves differently for the interval $-1/2 < w_e < 1/3$ than for
$w_e \le -1/2$, where the interaction modifies the DE evolution.
In the matter era, a non-adiabatic blow-up may thus happen if
 \be
-1 < w_e \le -1/2\,.
 \ee
A detailed analysis shows however that in the interval
 \be
-2/3 < w_e < -1/2\,,
 \ee
the ``blow-up'' mode is in fact a decaying mode, and hence
(non-standard) adiabatic evolution on super-Hubble scales is
possible. In the interval $-1 < w_e < -2/3$ the non-adiabatic
``blow-up'' mode is rapidly increasing, and will dominate unless
$|\Gamma|$ is suitably small. Therefore, a blow-up in the matter
era will make large interaction models with $-1 < w_e < -2/3$
non-viable, while the blow-up in the radiation era ruins all
interacting models (no matter how weak) with $-1 < w_e < -4/5$. We
summarize these results in Table~\ref{table:summary}.

\begin{table*}
\caption{The evolution of perturbations on super-Hubble scales
with various values of the early dark energy equation of state parameter
in the radiation and matter dominated eras (RD and MD respectively). ``Adiabatic'' means that
it is possible to specify adiabatic initial conditions so that the
total gauge invariant curvature perturbation $\zeta$ stays
constant on super-Hubble scales. ``Adiabatic (standard)'' means
that the behaviour of perturbations at early times on super-Hubble
scales is the same as in the non-interacting
model.\label{table:summary}}
  \begin{tabular}{|l|l|l|l|}
  \hline
$w_{de}$ in the RD or MD era & Radiation dominated era (RD) &
Matter dominated era (MD) & Viable?
  \\&&& \\
 \hline
$w_{de}<-1$ & adiabatic & adiabatic & viable, but phantom \\
$-1< w_{de} < -4/5 $ & ``blow-up'' isocurvature growth & ``blow-up''
isocurvature growth & non-viable \\   $ -4/5 \le w_{de} < -2/3$ & adiabatic &
isocurvature growth & viable, if $\Gamma$ small enough \\
$ -2/3 \le w_{de} < -1/2$ &  adiabatic (standard) &  adiabatic & viable \\
$ -1/2 \le w_{de} < +1/3 $ & adiabatic (standard) &  adiabatic (standard)
 & viable \\
\hline
  \end{tabular}
\end{table*}

%\section{Super-horizon initial conditions}

Assuming tight coupling between photons and baryons, so that $V_b
= V_\gamma$, passing from conformal time $\tau$ to the time
variable $x = k\tau $, using a rescaled velocity $\tilde V_A =
V_A/x$ and a rescaled anisotropic stress $\tilde \Pi_A =
\Pi_A/x^2$, as in \citet{Doran:2003xq}, we obtain the following
evolution equations:
% comment removed
\bea    \frac{d \Delta_{c}}{d\ln x}& =& -x^2\tilde{V_c} -
\frac{\Gamma}{\h_0}\, \alpha \, x^2\left(\frac{3}{2}
(1+w)\tilde{V} + 2 \Omega_\nu \tilde{\Pi}_\nu   + 2\Psi \right)\,,
\label{eq.Dc_newnew}\\     \frac{d \tilde{V}_c}{d\ln x} & =& -2
\tilde{V_c} + \Psi\,, \label{eq.Vc_newnew}\\     \frac{d \Delta_{\gamma}}{d\ln x}& =&
-\frac{4}{3} x^2 \tilde V_{\gamma}\,, \label{eq.Dg_newnew}\\
\frac{d \tilde{V}_{\gamma}}{d\ln x} &=&  \frac{1}{4}
\Delta_{\gamma}- \tilde V_{\gamma}  + \Omega_\nu \tilde{\Pi}_\nu +
2\Psi \,,
\label{eq.Vg_newnew}\\
\frac{d \Delta_b}{d\ln x} &=& - x^2 \tilde{V}_\gamma \,,
\label{eq.Db_newnew}\\     \frac{d \Delta_{\nu} }{d\ln x}&=&
-\frac{4}{3} x^2 \tilde{V}_\nu \,,\label{eq.Dn_newnew}\\   \frac{d
\tilde{V}_{\nu}}{d\ln x} &=& \frac{1}{4}\Delta_\nu -\tilde{V}_\nu
-\frac{1}{6}x^2 \tilde{\Pi}_\nu +  \Omega_\nu \tilde{\Pi}_\nu + 2
\Psi  \,,\\     \frac{d \tilde{\Pi}_{\nu}}{d\ln x}
 &=& \frac{8}{5} \tilde{V}_\nu - 2 \tilde{\Pi}_\nu \,,
 \label{eq.Pnu_newnew} \\
\frac{d \Delta_{de}}{d\ln x} &=& 3(w_e-1)\left\{ \Delta_{de} + 3
(1+w_e)\left[ -\Psi -\Omega_\nu \tilde{\Pi}_\nu  \right]
+(1+w_e)\left[ 3 - \frac{x^2}{3(w_e -1)} \right]
\tilde{V}_{de} \right\} \nonumber \\
  && +
\frac{\Gamma}{\h_0}\alpha x^2 \frac{\rho_c}{\rho_{de}} \left[
\Delta_c -\Delta_x + 3 (1-w_e) \tilde{V}_{de} - ( 3 w_e -5 )
\left[ \Psi + \Omega_\nu \tilde{\Pi}_\nu \right] + \frac{3}{2}
(1+w)\tilde{V}  \right]
\label{eq.Dx_newnew} \\
\frac{d \tilde{V}_{de}}{d\ln x}&=& \frac{\Delta_{de}}{1+w_e} +
\tilde{V}_{de} +  3 \Omega_\nu \tilde{\Pi}_\nu + 4\Psi  +
\frac{\Gamma}{\h_0}\alpha x^2 \frac{\rho_c}{\rho_{de}}
\frac{\tilde{V}_c - 2 \tilde{V}_{de} - \Omega_\nu \tilde{\Pi}_\nu
- \Psi }{1+w_e}
    \label{eq.Vx_newnew} \,.
 \eea
Here $\alpha = (\h_0/k)^2\sqrt{\Omega_{r0}}$, and we used the
Einstein equations
 \bea
\Phi &=& \frac{3}{2} x^{-2} \left\{ \Delta + 3 (1+w) \left[
\tilde{V}
 -\Phi \right] \right\}\,,\label{eq.Phi1}\\
\frac{d\Phi}{d\ln x}  &=&  \Psi - \frac{3}{2} (1+w) \tilde{V} \,,
\label{eq.dPhi1}\\
\Phi &=& -\Psi - 3 w \tilde \Pi \,,\label{eq.anis.stress1}
 \eea
to eliminate $\Phi$ in favour of $\Psi$.

Using Eqs.~(\ref{eq.Phi1}) and (\ref{eq.anis.stress1}), we have
 \be
\Psi = -\frac{\sum_{A = c, b, \gamma, \nu, de} \Omega_A \left[
\Delta_A + 3 \,(1+w_A) \tilde{V}_A \right]}{\sum_{A = c, b,
\gamma, \nu, de} 3 \,(1+w_A) \Omega_A + \frac{2}{3} x^2} -
\Omega_\nu \tilde{\Pi}_\nu \,. \label{eq.Psi}
 \ee
The total velocity appearing in Eqs.~(\ref{eq.Dc_newnew}) and
(\ref{eq.Dx_newnew}) is
 \be \label{eq.V}
\tilde{V} (1+w) = \sum_{A = c, b, \gamma, \nu, de} \Omega_A
(1+w_A) \tilde{V}_A\,.
 \ee
Recalling that $\rho = \sum \rho_A$, $(1+w) = \sum \Omega_A
(1+w_A)$, $\Delta = \sum \Omega_A \Delta_A$, and $\tilde\Pi = \sum
\Omega_A \tilde\Pi_A = \Omega_\nu \tilde\Pi_\nu$, we then see that
Eqs.~(\ref{eq.Dc_newnew}--\ref{eq.Vx_newnew}) form a set of 10
linear differential equations for 10 perturbation variables
$\Delta_A$, $\tilde V_A$ and $\tilde\Pi_\nu$. (Note that $\Pi_A=0$
for $A\neq\nu$.)

Since we are interested in the early radiation era, we make the
approximation $\Omega_A = \rho_A / \rho \simeq \rho_A /\rho_r$.
Using Eqs.~(\ref{eq:confHatRD}), (\ref{eq:blowuprevol}),
(\ref{eq:rhocorhor}) and (\ref{eq.omega_de_evol}), and the
(standard non-interacting) background evolution of photons,
baryons and neutrinos, we obtain
 \bea
 && \Omega_b = \frac{\rho_b}{\rho_r} =
\frac{\Omega_{b0}}{\Omega_{r0}} \, a =
\frac{\Omega_{b0}}{\sqrt{\Omega_{r0}}}\frac{\h_0}{k} \, x =
\omega_1 \,x \,, \quad \quad \quad \Omega_c =
\omega_2 \, x \,, \nonumber\\
&& \Omega_{de} = \tilde C\omega_2 x^3\ \ \mbox{for } w_e \le
-2/3,\quad \quad \quad \mbox{ and }\quad \quad \quad \Omega_{de} =  C\omega_2 x^{1-3 w_e}\
\ \mbox{for } {-2}/3 < w_e < 1/3\,, \nonumber \\
&& \Omega_\nu = \rho_\nu /\rho_r = R_\nu \,, \quad \quad \quad
\Omega_\gamma = 1- \Omega_b - \Omega_c - \Omega_{de} -\Omega_\nu
\,. \label{eq.Omegas_expl}
 \eea

The next step of the method proposed in \citet{Doran:2003xq}
consists in writing the system of differential equations
(\ref{eq.Dc_newnew}--\ref{eq.Vx_newnew}) in a matrix form:
\be
\label{eq.dUdlnx_gen} \frac{d \bmath{U}}{d \ln x} = \textbfss{A}(x) \bmath{U}(x)
\,, \ee where \be \bmath{U}^T = \left\{ \Delta_c, \, \tilde{V}_c, \,
\Delta_\gamma, \, \tilde{V}_\gamma, \, \Delta_b, \, \Delta_\nu, \,
\tilde{V}_\nu ,\, \tilde{\Pi}_\nu , \, \Delta_{de}, \,
\tilde{V}_{de} \right\}\,,
 \ee
and the matrix $\textbfss{A}(x)$ can be read from Eqs.
(\ref{eq.Dc_newnew}--\ref{eq.Vx_newnew}) after substituting Eqs.
(\ref{eq.Psi}--\ref{eq.Omegas_expl}) and the background evolution
of $\rho_c/\rho_{de}$ from (\ref{eq:blowuprevol}) or
(\ref{eq.r_evol}), depending on the value of $w_e$.

The initial conditions are specified for modes well outside the
horizon, i.e. for $x \ll 1$.  There will be several independent
solution vectors to Eq. \ref{eq.dUdlnx_gen}, that we write as $x^{\lambda_i}\bmath{U}^{(i)}$. If no term of $\textbfss{A}(x)$
diverges for $x\to 0$, then we can approximate $\textbfss{A}$ by a constant
matrix $\textbfss{A}_0 = \lim_{x\to0} \textbfss{A}(x)$. If we require more accuracy we
can expand $\textbfss{A}(x)$ up to a desired order in $x$. For example, up to
order $x^3$ the matrix $\textbfss{A}(x)$ contains the constant term $\textbfss{A}_0$ as
well as terms proportional to $x$, $x^2$, and $x^3$ in the case
where $w_e \le -2/3$. However, in the case $-2/3 < w_e \le 1/3$,
$\textbfss{A}(x)$ contains in addition to integer powers of $x$ also
non-integer powers $1-3w_e$, $2-3w_e$, $3-3w_e$, etc., and
$2+3w_e$, $3+3w_e$, $4+3w_e$, etc. The listed ones and possibly
their multiples can fall in the range $(0,3)$. For a given $w_e$,
however, one should drop those which turn out to be higher order
than $x^3$.

Thus going beyond zeroth order, up to order $x^3$, we can expand
$\textbfss{A}$ and each solution $x^{\lambda_i}\bmath{U}^{(i)}$ as
 \bea
\textbfss{A}(x) &\simeq& \textbfss{A}_0 + \textbfss{A}_1\, x + \textbfss{A}_2\, x^2 + \textbfss{A}_3\, x^3 + \sum_{j=0}^3
\left[ \sum_{n=1}^N \left( \textbfss{B}_{nj} \, x^{n(1-3w_e)+j} \right) +
\textbfss{C}_{j}\,x^{2+3w_e + j} \right]\,,
\label{eq:Axexpansion}\\
x^{\lambda_i} \bmath{U}^{(i)}(x) &\simeq&  x^{\lambda_i} \left\{
\bmath{U}_0^{(i)} + \bmath{U}_1^{(i)}\, x + \bmath{U}_2^{(i)}\, x^2
+\bmath{U}_3^{(i)}\, x^3 + \sum_{j=0}^3 \left[ \sum_{n=1}^N \left(
\bmath{U}_{Bnj}^{(i)} x^{n(1-3w_e)+j}  \right) + \bmath{U}_{Cj}^{(i)}
x^{2+3w_e + j}\right]\right\},
\label{eq:Uxexpansion}
 \eea
where $\textbfss{A}_j$, $\textbfss{B}_{nj}$, and $\textbfss{C}_{j}$ are constant (not depending on
the time variable $x$) matrices, and $\bmath{U}_{j}^{(i)}$ are
constant vectors. Note that for $w_e \le -2/3$ all $\textbfss{B}_{nj}$,
$\textbfss{C}_{j}$, $\bmath{U}_{Bnj}^{(i)}$, and $\bmath{U}_{Cj}^{(i)}$ terms
vanish. For simplicity, we demonstrate below this case,
which leads to only integer powers. Substituting
Eqs.~(\ref{eq:Axexpansion}) and (\ref{eq:Uxexpansion}) into the
evolution equation (\ref{eq.dUdlnx_gen}) and equating order by
order, we obtain
 \bea
\textbfss{A}_0\bmath{U}_0^{(i)} &=& \lambda_i \bmath{U}_0^{(i)}\,,\ \mbox{i.e.
$\lambda_i$ is an eigenvalue of $\textbfss{A}_0$, and $\bmath{U}_0^{(i)}$
is an eigenvector of $\textbfss{A}_0$},\\
\bmath{U}_1^{(i)} &=& -\left[\textbfss{A}_0 -
(\lambda_i+1)\mathds{1}\right]^{-1} \left[ \textbfss{A}_1 \bmath{U}_0^{(i)} +
\textbfss{A}_0 \bmath{U}_1^{(i)} \right]\,,
\label{eq:U1}\\
\bmath{U}_2^{(i)} &=& -\left[\textbfss{A}_0 -
(\lambda_i+2)\mathds{1}\right]^{-1} \left[ \textbfss{A}_2 \bmath{U}_0^{(i)} +\textbfss{A}_1
\bmath{U}_1^{(i)} +\textbfss{A}_0 \bmath{U}_2^{(i)}
\right]\,,\\
\bmath{U}_3^{(i)} &=& -\left[\textbfss{A}_0 -
(\lambda_i+3)\mathds{1}\right]^{-1} \left[ \textbfss{A}_3 \bmath{U}_0^{(i)} +\textbfss{A}_2
\bmath{U}_1^{(i)} + \textbfss{A}_1 \bmath{U}_2^{(i)} + \textbfss{A}_0 \bmath{U}_3^{(i)}
\right]\,.\label{eq:U3}
 \eea

Now the general solution to the differential equation
(\ref{eq.dUdlnx_gen}) is a linear combination of solutions
$x^{\lambda_i}\bmath{U}^{(i)}$: \be \bmath{U}(x) =  \sum_i c_i \,
x^{\lambda_i}\, \bmath{U}^{(i)}(x)\,, \ee where $c_i$ are dimensionless
constants. If we define an initial reference time $t_{\rm init}$,
then the constants $\tilde c_i = c_i x_{\rm init}^{\lambda_i}$
represent the initial contribution of the vector $\bmath{U}^{(i)}$ to
the total perturbation vector $\bmath{U}(x_{\rm init})$.  The imaginary
part of $\lambda_i$ represents oscillations of
$x^{\lambda_i}\bmath{U}_0^{(i)}(x)$, while the real part gives its
power-law behaviour: $x^{\lambda_i}\bmath{U}_0^{(i)}(x) =  x^{{\rm
Re}(\lambda_i)} \cos[{\rm Im}(\lambda_i)\,\ln x]$. 
%A positive
%${\rm Re}(\lambda_i)$ means a growing solution; the larger ${\rm
%Re}(\lambda_i)$, the faster is the growth of all perturbation
%variables.
%%%%%%%%%%%%%%%%%%%
\comment{$\textbfss{A}(x) \simeq \textbfss{A}_0$. We find that this is the case for
$-2/3 < w_e < 1/3$. At zeroth order in $x$, Eq.
(\ref{eq.dUdlnx_gen}) becomes \be  \label{eq.dUdlnx_app0} \frac{d
\bmath{U}}{d \ln x} = \textbfss{A}_0 \, \bmath{U} \,, \ee and the solution is \be
\label{eq.U_app0} \bmath{U}(x) = \sum_i c_i \left(
\frac{x}{x_i}\right)^{\lambda_i}\, \bmath{U}^{(i)} \,, \ee where
$\bmath{U}^{(i)}$ and $\lambda_i$ are the eigenvectors and
eigenvalues of $\textbfss{A}_0$, respectively, and where the coefficients
$c_i$ represent the initial contribution of the eigenvector
$\bmath{U}^{(i)}$ to the total perturbation $\bmath{U}(x)$. }
%%%%%%%%%%%%%%%%%%%
The contribution corresponding to the eigenvalue(s) with largest
real part, ${\rm Re}(\lambda_i)$, will dominate as time goes by,
while initial contributions from eigenvectors corresponding to
$\lambda_i$ with smaller real part will become negligible compared
to the dominant mode. Hence, to set initial conditions deep in the
radiation era but well after inflation, it is sufficient to
specify the contribution coming from  mode(s) with largest ${\rm
Re} (\lambda_i)$.

From now on we divide the treatment into two cases $-2/3 < w_e <
1/3$ and $w_e \le -2/3$. Before proceeding, we should point out
that the matrix method presented in \citet{Doran:2003xq} and
applied to non-interacting constant-$w_{de}$ dark energy,
represents a systematic and efficient approach for finding initial
conditions. Once the matrix $\textbfss{A}(x)$ has been read from the set of
first order differential equations
(\ref{eq.Dc_newnew}--\ref{eq.Vx_newnew}), one can feed it into a
symbolic mathematical program such as Maple or Mathematica and
easily extract the constant part $\textbfss{A}_0$ as well as the other parts
(such as $\textbfss{A}_1$, $\textbfss{A}_2$, ..., $\textbfss{B}_{nj}$, $\textbfss{C}_{j}$) up to any desired
order. Then it is simple linear algebra to find the eigenvalues
$\lambda_i$ and eigenvectors $\bmath{U}_0^{(i)}$ of $\textbfss{A}_0$ and, if
higher order solutions in $k\tau$ are needed, to substitute these
step by step into Eqs. (\ref{eq:U1})--(\ref{eq:U3}) etc. in order
to find the solutions $x^{\lambda_i}\bmath{U}^{(i)}(x)$.

\subsection{Case $-2/3 < w_e < 1/3$}

We substitute $\Psi$ from Eq.~(\ref{eq.Psi}), $\tilde V$ from
Eq.~(\ref{eq.V}), and the energy density parameters from
Eq.~(\ref{eq.Omegas_expl}) into
Eqs.~(\ref{eq.Dc_newnew}--\ref{eq.Vx_newnew}). Then, using
Eq.~(\ref{eq.r_evol}) for
$\rho_c/\rho_{de}$ in the last two of
them, and taking the limit $x\to 0$, we find the $\textbfss{A}_0$ matrix:

\be 
\label{eq:A0forw>-2/3}
\textbfss{A}_0= 
\left(
\begin{array}{cccccccccc}
 0 & 0 & 0 & 0 & 0 & 0 & 0 & 0 & 0 & 0 \\
&&&&&&&&&\\ 0 & -2 & \frac{\mathcal{N}}{4} & \mathcal{N} & 0 &
-\frac{R_\nu}{4} & -R_\nu & -R_\nu & 0 & 0 \\ &&&&&&&&&\\
 0 & 0 & 0 & 0 & 0 & 0 & 0 & 0 & 0 & 0 \\
&&&&&&&&&\\ 0 & 0 & \frac{2R_\nu -1}{4} & 2R_\nu -3 & 0 &
-\frac{R_\nu}{2} & -2R_\nu & -R_\nu & 0 & 0 \\ &&&&&&&&&\\ 0 & 0 &
0 & 0 & 0 & 0 & 0 & 0 & 0 & 0 \\ &&&&&&&&&\\ 0 & 0 & 0 & 0 & 0 & 0
& 0 & 0 & 0 & 0 \\ &&&&&&&&&\\ 0 & 0 & \frac{\mathcal{N}}{2} & 2
\mathcal{N}  & 0 & \frac{1 - 2R_\nu}{4} &  -1 - 2R_\nu & -R_\nu &
0 & 0 \\ &&&&&&&&&\\ 0 & 0 & 0 & 0 & 0 & 0 & \frac{8}{5} & -2 & 0
& 0 \\ &&&&&&&&&\\ 0 & 0 & \frac{9 \mathcal{N} \left({w_e}^2 -1
\right) }{4} &  9 \mathcal{N} \left( {w_e}^2 -1 \right)  & 0 &
\frac{9R_\nu \left(1- {w_e}^2 \right) }{4} &
   9R_\nu\left(1 - {w_e}^2 \right)  & 0 &  3\left(w_e
   -1\right)
   & 9\left({w_e}^2 -1 \right) \\
&&&&&&&&&\\ 0 & 0 & \mathcal{N} & 4 \mathcal{N}  & 0 & -R_\nu &
-4R_\nu & -R_\nu & \frac{1}{1 + w_e} & 1
\end{array}
\right) 
\ee 
where $\mathcal{N} = R_\nu -1$.

The eigenvalues of $\textbfss{A}_0$ are \be \lambda_i = \left\{
-2,-1,0,0,0,0,- \frac{5}{2}   - \frac{\sqrt{1 - 32\,R_\nu
/5}}{2},- \frac{5}{2} + \frac{\sqrt{1 -  32\,R_\nu /5}}{2},
\lambda_{\rm{d}}^{-}, \lambda_{\rm{d}}^{+} \right\} \,,
\label{eq.eigenvals_w>-2/3} \ee where \be \lambda_{\rm{d}}^{\pm} =
\frac{-2 + 3\,w_e}{2} \pm \frac{{\sqrt{-20 + 12\,w_e +
9\,{w_e}^2}}}{2} \,. \label{eq:lambdasforw>-2/3} \ee For the range
$0<R_\nu<0.405$ and $-2/3 < w_e < 1/3 $, all eigenvalues have a
non-positive real part. In (\ref{eq:lambdasforw>-2/3}) the term
inside the square root, $-20 + 12\,w_e + 9\,{w_e}^2$, falls
between $-24$ and $-15$, and hence ${\rm
Re}(\lambda_{\rm{d}}^{\pm}) = -1 + 3\,w_e/2$, falls between $-2$
and $-1/2$.

As explained in \citet{Doran:2003xq}, since the eigenvalue with
largest real part, $\lambda = 0$, is fourfold degenerate, it is
possible to choose a basis from the subspace spanned by the
eigenvectors with eigenvalue $\lambda = 0$, so that physically
meaningful choices can be made for the initial condition vector. One can form 4 independent linear
combinations from the four vectors with $\lambda =0$. The physical
choices are an adiabatic mode and 3 isocurvature modes. Here we
choose adiabatic initial conditions, specified by the condition
that the gauge-invariant entropy perturbations $S_{AB}$ of every
$A,\, B\, = \gamma, \, \nu,\, c,\,b$ vanish, where
\citep{Malik:2002jb} \be \label{eq.entropy} S_{AB} = -
3\h\frac{\rho_A}{\rho'_A} \Delta_A + 3\h \frac{\rho_B}{\rho'_B}
\Delta_B\,. \ee We will show later the interesting new result that
demanding adiabaticity between the standard constituents
automatically guarantees adiabaticity with respect to DE.

We should remind the reader that for the interacting constituents
the coupling appears in the continuity equation, and we should not
use blindly the standard result  \be \label{eq.entropy_w>-2/3}
S_{AB} =  \frac{\Delta_A}{(1+w_A)} - \frac{\Delta_B}{(1+w_B)}\,,
\ee where the $1+w_A$ factors result from applying the continuity
equation to $\rho'_A/\rho_A$. Indeed, for cold dark matter in the early
radiation era we find, using Eqs.(\ref{eq.cont_rhoc}),
(\ref{eq:ourcoupling}), and (\ref{eq:confHatRD}),
 \be
- 3\h\frac{\rho_c}{\rho'_c} \Delta_c =
\frac{\Delta_c}{(1+w_c)+({H_0\sqrt{\Omega_{r0}}}/{3k^2})(k\tau)^2}\,,
 \ee
where $w_c=0$. For DE, we find using Eqs.(\ref{eq.cont_rhode}),
(\ref{eq:ourcoupling}), (\ref{eq:confHatRD}), and
(\ref{eq.r_evol})
 \be
- 3\h\frac{\rho_{de}}{\rho'_{de}} \Delta_{de} =
\frac{\Delta_{de}}{(1+w_{de})-({H_0\sqrt{\Omega_{r0}}}/
{3k^2}{C})(k\tau)^{2+w_e}}\,.
 \ee
At zeroth order in $x=k\tau$ we incidentally regain the standard
non-interacting result (\ref{eq.entropy_w>-2/3}). From $S_{AB} =
0$ it then follows that
 \be \label{eq.adiabatic_condition}
\Delta_c = \Delta_b = \frac{3}{4} \Delta_\gamma = \frac{3}{4}
\Delta_\nu \,.
 \ee
Imposing this condition on a linear combination of the four
eigenvectors with eigenvalue $\lambda = 0$ we obtain
\begin{equation}
\label{eq.adiabatic_mode} \bmath{U}^{({\rm adi}) }_0 = \left(
\begin{array}{c}
\Delta_c \\ \tilde V_c\\ \Delta_{\gamma} \\ \tilde V_{\gamma}
\\ \Delta_b \\ \Delta_{\nu} \\ \tilde V_{\nu} \\ \tilde \Pi_{\nu}
\\ \Delta_{de} \\
\tilde{V}_{de}
\end{array}
\right)_{\mbox{adiabatic}} = C_1 \left(
\begin{array}{c}
3/4  \\
-(5/4) \mathcal{P} \\
1 \\
- (5/4) \mathcal{P}\\
3/4  \\
1 \\
- (5/4) \mathcal{P}\\
- \mathcal{P} \\
(3/4) \left( 1 + w_e \right) \\
- (5/4) \mathcal{P}
\end{array}
\right) \,,
\end{equation}
where $\mathcal{P} = \left( 15 + 4\,R_\nu \right)^{-1}$, and $C_1$
is a dimensionless normalization constant corresponding to, e.g.,
$\Delta_\gamma$ and $\Delta_\nu$ at the initial time.

The vector (\ref{eq.adiabatic_mode}) is identical to the standard
adiabatic initial condition vector (see \citet{Doran:2003xq}). In
particular, it should be noticed that although we did not require
adiabaticity of DE, (\ref{eq.adiabatic_mode}) automatically
satisfies the condition $S_{de,A} =0$ for all $A =\gamma, \,
\nu,\, c,\,b$. In \citet{Doran:2003xq} this result was found for
non-interacting dark energy. Here we have now shown that also
interacting dark energy is automatically adiabatic, once cold dark
matter, baryons, photons and neutrinos are set to be adiabatic.

Finally, since all components of the vector
(\ref{eq.adiabatic_mode}) are different from zero, it is not
necessary to compute terms to higher order in $x$, and we use Eq.
(\ref{eq.adiabatic_mode}) as our adiabatic initial condition for
the computation of the CMB power spectrum for models with $-2/3 <
w_e < 1/3$.

\citet{Lee:2006za} have reported that the
quintessence isocurvature mode decays away
(in an interacting quintessence model which is quite similar to our set-up). 
After our systematic derivation of initial conditions,
this decay can be tracked down to the fact that 
${\rm Re}(\lambda_{\rm{d}}^{\pm})$ 
in Eq.~(\ref{eq.eigenvals_w>-2/3})
are negative. The reason for this is that
in quintessence models the early-time equation of state parameter
is typically larger than $-2/3$, indeed positive [but 
in  \citet{Lee:2006za} less than $+2/3$]. So 
${\rm Re}(\lambda_{\rm{d}}^{\pm})$ is negative, and hence
the isocurvature mode decays. In the next subsection we will
see that also in the range $-4/5 \le w_e \le -2/3$ (or $w_e < -1$)
the DE isocurvature mode decays (although in this case 
the interaction affects the evolution of DE perturbations),
whereas in the range $-1 < w_e < -4/5$ the DE isocurvature mode
is a rapidly growing mode as recently realised
by \cite{Valiviita:2008iv}. 

\subsection{Case $w_e \le -2/3$}

Since at early times the equation of state can be approximated as
a constant, $w_e = w_0 +w_a$, this case has already been studied
in \citet{Valiviita:2008iv}, where a constant $-1< w_{de} \le -2/3$
was analysed. A serious non-adiabatic large-scale instability that
excludes these models was found whenever $-1 < w_{de} < -4/5$, no
matter how weak the interaction was. However, we notice that there
is a limited region of parameter space, $-4/5 \le w_{de} \le
-2/3$, where the instability can possibly be avoided. In the case
of a constant DE equation of state parameter this range would be
observationally disfavoured, since for example supernova data
require that $w_{de}$ is closer to $-1$ at recent times. In the
case of time varying $w_{de}(a)$ we do not have this problem as
$w_0$ can be close to $-1$ while $-4/5 < w_e  \le -2/3$. In the
following we repeat the analysis of initial conditions done in
\citet{Valiviita:2008iv}, but using the matrix method of
\citet{Doran:2003xq}, extended to include the interaction, and give the conditions for a viable
cosmology.

Substituting $\Psi$ from Eq.~(\ref{eq.Psi}), $\tilde V$ from
Eq.~(\ref{eq.V}), and the energy density parameters from
Eq.~(\ref{eq.Omegas_expl}) into
Eqs.~(\ref{eq.Dc_newnew}--\ref{eq.Vx_newnew}), as well as
$\rho_c/\rho_{de}$ from Eq.~(\ref{eq:blowuprevol}) into the last
two of them, and taking the limit $x\to 0$, we find the $\textbfss{A}_0$
matrix, which is very similar to our previous result,
Eq.~(\ref{eq:A0forw>-2/3}). This happens because everything 
remains unchanged, except that we must replace in
Eqs.~(\ref{eq.Dx_newnew}) and (\ref{eq.Vx_newnew}) the evolution
of $\rho_c/\rho_{de}$ with Eq.~(\ref{eq:blowuprevol}), and
whenever $\Omega_{de}$ appears we must now substitute the $\propto
x^3$ behaviour from (\ref{eq.Omegas_expl}), instead of the $\propto
x^{1-3 w_e}$ behaviour. Therefore only the last two rows in
(\ref{eq:A0forw>-2/3}) are modified, and will now read
%which is presented in the Appendix \ref{App:A-matices},
%Eq.~(\ref{eq:A0forw<-2/3}). Note that the matrix is almost the
%same as in the case $-2/3 < w_e < 1/3$, except that we have
%replaced in Eqs.~(\ref{eq.Dx_newnew}) and (\ref{eq.Vx_newnew}) the
%evolution of $\rho_c/\rho_{de}$ with Eq.~(\ref{eq:blowuprevol}),
%and whenever $\Omega_{de}$ appeared we have now substituted
%$\propto x^3$ behaviour from (\ref{eq.Omegas_expl}) instead of
%$\propto x^{1-3 w_e}$ behaviour.
%\begin{minipage}{200mm}
%\begin{center}
\be \label{eq:A0forw<-2/3}
%\textbfss{A}_0=
\left(
\begin{array}{cccccccccc}
%0 & 0 & 0 & 0 & 0 & 0 & 0 & 0 & 0 & 0 \\
%&&&&&&&&&\\
%0 & -2 & \frac{\mathcal{N}}{4} & \mathcal{N}& 0 & -\frac{R_\nu}{4} & -R_\nu & -R_\nu & 0 & 0 \\
%&&&&&&&&&\\
% 0 & 0 & 0 & 0 & 0 & 0 & 0 & 0 & 0 & 0 \\
%&&&&&&&&&\\
% 0 & 0 & \frac{2 R_\nu -1}{4} & 2 R_\nu -3 & 0 & -\frac{R_\nu}{2} & -2 R_\nu & -R_\nu & 0 & 0 \\
%&&&&&&&&&\\
% 0 & 0 & 0 & 0 & 0 & 0 & 0 & 0 & 0 & 0\\
%&&&&&&&&&\\
% 0 & 0 & 0 & 0 & 0 & 0 & 0 & 0 & 0 & 0\\
%&&&&&&&&&\\
% 0 & 0 & \frac{\mathcal{N}}{2} & 2 \mathcal{N}  & 0 & \frac{1 -
%2 R_\nu}{4} &   -(1 +2 R_\nu) & -R_\nu & 0 & 0\\
%&&&&&&&&&\\
%0 & 0 & 0 & 0 & 0 & 0 & \frac{8}{5} & -2 & 0 & 0\\
%&&&&&&&&&\\
2 + 3 w_e & 0 & \frac{\mathcal{N}
 \left( 1 + 9 w_e \right) }{4} &
   3 \mathcal{N}  \left(w_e -1
\right)  & 0 & -\frac{R_\nu \left( 1 + 9 w_e \right)  }{4} &
   -3 R_\nu \left(w_e -1\right)  & 0 & -5 & 3 \left( w_e -1\right) \\
&&&&&&&&&\\
 0 & \frac{2 + 3 w_e}{1 + w_e} & \frac{\mathcal{N}  \left( 2 + w_e \right) }
    {4 \left( 1 + w_e \right) } & \frac{\mathcal{N}  \left( 2 + w_e \right) }{1 +
w_e} & 0 &
   -\frac{R_\nu \left( 2 + w_e \right)
 }{4 \left( 1 + w_e \right) } &
   -\frac{R_\nu \left( 2 + w_e \right) }{1 +
w_e}  & -R_\nu & \frac{1}{1 + w_e} &
   -\frac{3 +5 w_e}{1 + w_e}
\end{array}
\right). \ee
%\end{center}
%\end{minipage}
%
The eigenvalues of $\textbfss{A}_0$ are
 \be \label{eq.eigenvals_w<-2/3}
\lambda_i =\left\{-2,-1,0,0,0,0,-\frac{5}{2} - \frac{\sqrt{1-32
R_\nu/5}}{2}, -\frac{5}{2} + \frac{\sqrt{1-32 R_\nu/5}}{2},
\lambda_{\rm g}^-, \lambda_{\rm g}^+\right\}\,,
 \ee
where
 \be
\lambda_{\rm g}^{\pm} \equiv \frac{-5 w_e- 4 \pm \sqrt{3
w_e^2-2}}{1+ w_e} \,.
 \ee
The first eight eigenvalues coincide with the previous case,
Eq.~(\ref{eq.eigenvals_w>-2/3}). Of those, four have a negative
real part, corresponding thus to modes that will decay away
quickly, and that we can neglect. The last two eigenvalues,
$\lambda_{\rm g}^{\pm}$, are instead very different from the
previous case, and depend on the value of $w_e$. The eigenvalue with a largest real part, $\lambda_{\rm
g}^+$, is real and positive for $-1 < w_e \le -\sqrt{2/3}$. In
addition to this, ${\rm Re}(\lambda_{\rm g}^+)$ is positive also
in the small range $-\sqrt{2/3} <  w_e < -4/5$. Therefore ${\rm
Re}(\lambda_{\rm g}^+) > 0$ for $-1 < w_e < -4/5$. This
corresponds to the blow-up solution found in
\citet{Valiviita:2008iv}; $\lambda_{\rm g}^+$ is larger, the closer
$w_e$ is to $-1$. There is no blow-up of perturbations for $-4/5
\le w_e \le -2/3$, because then the largest ${\rm Re}(\lambda_i)$
are zero.

\subsubsection{Case $-1 < w_e < -4/5$; non-adiabatic blow-up}

We now calculate the initial condition vector $\bmath{U}^{(\rm g)}
(x)$ corresponding to the fastest growing mode, $\lambda_{\rm
g}^+$. At zeroth order in $x$, it is given by \be
\label{eq.buvect-0thorder} \bmath{U}^{({\rm g}) T}_0(x) = \left\{
0,\,0,\,0,\,0,\,0,\,0,\,0,\,0,\,-1+\sqrt{3 w_e^2 -2},\,1 \right\}
\,. \ee In this case, since only the last
two components of the vector are different from zero, we need to
compute higher order corrections. It turns out that an expansion up to
$x^3$ is necessary, and as explained both before and after
Eqs.~(\ref{eq:Axexpansion}) and (\ref{eq:Uxexpansion}), the
expansion contains only integer powers of $x$, when $ w_e \le
-2/3$. Therefore we have \bea
\textbfss{A}(x) &\simeq& \textbfss{A}_0 + \textbfss{A}_1\, x + \textbfss{A}_2\, x^2 + \textbfss{A}_3\, x^3\,, \\
\bmath{U}^{(g)}(x) &\simeq& \bmath{U}_0^{(g)} + \bmath{U}_1^{(g)}\, x +
\bmath{U}_2^{(g)}\, x^2 +\bmath{U}_3^{(g)}\, x^3\,. \eea By substituting
$\textbfss{A}(x)$ and $x^{\lambda_{\rm g}^+}\bmath{U}^{(g)}(x)$ into the
evolution equation (\ref{eq.dUdlnx_gen}) and equating order by
order, we obtain
 \bea
\bmath{U}_1^{(g)} &=& - \left[ \textbfss{A}_0 - (\lambda_{\rm g}^+ +
1)\mathds{1}
\right]^{-1} \textbfss{A}_1 \bmath{U}_0\,,\\
\bmath{U}_2^{(g)} &=& - \left[ \textbfss{A}_0 - (\lambda_{\rm g}^+ +
2)\mathds{1}
\right]^{-1} (\textbfss{A}_2 \bmath{U}_0 +\textbfss{A}_1\bmath{U}_1)\,,\\
\bmath{U}_3^{(g)} &=& - \left[ \textbfss{A}_0 - (\lambda_{\rm g}^+ +
3)\mathds{1} \right]^{-1} (\textbfss{A}_3 \bmath{U}_0 +\textbfss{A}_2 \bmath{U}_1 + \textbfss{A}_1
\bmath{U}_2)\,.
 \eea
Using these formulas we find corrections to Eq.
(\ref{eq.buvect-0thorder}). Keeping for each perturbation variable
only the leading order (in $x$) terms, we obtain the following
initial condition vector:
 \be \label{eq.growing_mode}
\bmath{U}^{({\rm g})}(x)= \left(
\begin{array}{c}
\Delta_c   \phantom{\frac{\frac{\Gamma}{\h_0}}{w_e }} \\
\tilde V_c \phantom{\frac{\frac{\Gamma}{\h_0}}{w_e }}\\
\Delta_{\gamma} \phantom{\frac{\frac{\Gamma}{\h_0}}{w_e }} \\
\tilde V_{\gamma}\phantom{\frac{\frac{\Gamma}{\h_0}}{w_e }} \\
\Delta_b \phantom{\frac{\frac{\Gamma}{\h_0}}{w_e }} \\
\Delta_{\nu} \phantom{\frac{\frac{\Gamma}{\h_0}}{w_e }} \\
\tilde V_{\nu} \phantom{\frac{\frac{\Gamma}{\h_0}}{w_e }}\\
\tilde \Pi_{\nu}\phantom{\frac{\frac{\Gamma}{\h_0}}{w_e }} \\
\Delta_{de} \phantom{\frac{\frac{\Gamma}{\h_0}}{w_e }} \\
\tilde{V}_{de} \phantom{\frac{\frac{\Gamma}{\h_0}}{w_e }}
\end{array}\!\!\!\!\!\!\!\!\!
\right)_{\mbox{g}} = C_2 \left(
\begin{array}{c}
0  \phantom{\frac{\frac{\Gamma}{\h_0}}{w_e }}\\
\frac{\frac{\Gamma}{\h_0} \omega_3 {\left( 1 + w_e \right) }^3
    \left[ 8 R_\nu \left( 1 + w_e \right) \ncal -
      15 \left(  w_e -1 \right)  \left( 1 -w_e + \ncal
      \right)  \right]  x^3}{4
    \left(  w_e  -1 \right)  \left( 2 + 3 w_e \right)
    \mathcal{M} } \phantom{\frac{\frac{\Gamma}{\h_0}}{w_e }}
    \phantom{\frac{\frac{\Gamma}{\h_0}}{w_e }}\\
0 \phantom{\frac{\frac{\Gamma}{\h_0}}{w_e }}\\
\frac{-15 \frac{\Gamma}{\h_0} \omega_3 {\left( 1 + w_e \right) }^3
\left( 2 + \ncal \right)  x^3}
  {2 \left( 2 + 3 w_e \right)  \mathcal{M} }
  \phantom{\frac{\frac{\Gamma}{\h_0}}{w_e }}\\
0  \phantom{\frac{\frac{\Gamma}{\h_0}}{w_e }}\\
0 \phantom{\frac{\frac{\Gamma}{\h_0}}{w_e }}\\
\frac{-15 \frac{\Gamma}{\h_0} \omega_3 {\left( 1 + w_e \right) }^3
\left( 2 + \ncal \right)  x^3}
  {2 \left( 2 + 3 w_e \right)  \mathcal{M} }
  \phantom{\frac{\frac{\Gamma}{\h_0}}{w_e }}\\
\frac{4 \frac{\Gamma}{\h_0} \omega_3 {\left( 1 + w_e \right) }^3
    \left[ 3\left( 7 + 8 w_e + {w_e}^2 \right) +
    (5 + 3 w_e) \ncal  \right]
       x^3}{\left( 2 + 3 w_e \right)  \left( \ncal -
       2 w_e  \right)
 \mathcal{B} } \phantom{\frac{\frac{\Gamma}{\h_0}}{w_e }}\\
\ncal \phantom{\frac{\frac{\Gamma}{\h_0}}{w_e }}\\
1  \phantom{\frac{\frac{\Gamma}{\h_0}}{w_e }}
\end{array}\!\!\!\!\!\!\!\!\!\!\!\!\!\!\!\!\!\!\!\!\!
\right)\,, \ee 
where  
$\omega_3 = \omega_2 (H_0/k)^2 \sqrt{\Omega_{r0}} =
(H_0/k)^3\Omega_{r0}/a_{eq}$,  $\ncal$,  $\mathcal{M}$ and $\mathcal{B}$ are
$\ncal = \sqrt{3 {w_e}^2
-2}-1$,
 $\mathcal{M} = 5 \left[ 6 + 7 w_e + 3 {w_e}^3 + (3 + 5
w_e)\ncal
 \right] -4  R_\nu {\left( 1 + w_e \right) }^2
       \left( \ncal -1 - 3 w_e
\right)$ and
$\mathcal{B} =  8 R_\nu\left( 1 + w_e \right)^2 \left( 5
+ 3 w_e + 2 \ncal \right)  +
    5 \left\{ 9 \ncal -3 + w_e
        \left[ 13 + 14 \ncal + 3 w_e
           \left( 13 + 5 w_e + 3 \ncal \right)  \right]
           \right\} $.
This solution coincides with equations (63)--(70) of \citet{Valiviita:2008iv}, after substituting $n_\psi =
\lambda_{\rm g}^+ + \,3$, $J = 1 - 16 R_\nu  [\,5 (n_\psi +
2)(n_\psi + 1) \,+\, 8 R_\nu\,]^{-1}$, converting into Newtonian
gauge (by using Eqs. (\ref{eq.conversion}) with $B = E = 0$) and
conveniently renormalizing the vector. Equation
(\ref{eq.growing_mode}) is the initial condition vector for the
case $-1 < w_e \le -4/5$, when the dominant eigenvector is that
corresponding to $\lambda_{\rm g}^+$.

The initial condition (\ref{eq.growing_mode}) is trivially
adiabatic with respect to $\gamma$, $\nu$, $c$ and $b$, but not
with respect to DE. Indeed, for DE we find using
Eqs.(\ref{eq.cont_rhode}), (\ref{eq:ourcoupling}), and
(\ref{eq:blowuprevol}) \be \label{eq:departinS}
 - 3\h\frac{\rho_{de}}{\rho'_{de}} \Delta_{de}
= \frac{\Delta_{de}}{(1+w_{de})-(3w_{de}+2)/3} =  3\Delta_{de}\,.
\ee Therefore
 \be
S_{de\, A} = 3\Delta_{de} = C_2\ncal x^{\lambda_{\rm g}^+}\,,
 \ee
for any $A = \gamma$, $\nu$, $c$ or $b$. Even if we were able
to set the initial conditions at $\tau =0$ and demanded
adiabaticity there, after a short time the solution would not be
adiabatic. Thus Eq.~(\ref{eq.growing_mode}) represents the
non-adiabatic ``blow-up'' solution \citep{Valiviita:2008iv} for the
case $-1 < w_e < -4/5$.

\subsubsection{Cases $w_e < -1$ or $-4/5 \le w_e \le -2/3$;
adiabatic initial conditions}

In the range $-4/5 < w_e \le -2/3$, as well as for $ w_e < -1$, we
have ${\rm Re}(\lambda_{\rm g}^\pm)< 0$, so that the largest
eigenvalue is the fourfold degenerate $\lambda = 0$. [If $w_e =
-4/5$, then $\lambda=0$ is fourfold degenerate, and there are also two
oscillating solutions with ${\rm Re}(\lambda_{\rm g}^\pm)=0$.] We
look for a linear combination of the four eigenvectors
(corresponding to $\lambda=0$) that satisfies adiabaticity (see
Eq. (\ref{eq.adiabatic_condition})) of photons, neutrinos, baryons
and cold dark matter. The resulting eigenvector is
\begin{equation}\label{eq.adiabatic_mode_w<-2/3}
\bmath{U}^{({\rm adi}) }_0 = \left(
\begin{array}{c}
\Delta_c \\ \tilde V_c\\ \Delta_{\gamma} \\ \tilde V_{\gamma} \\
\Delta_b \\ \Delta_{\nu} \\ \tilde V_{\nu} \\ \tilde \Pi_{\nu} \\
\Delta_{de} \\
\tilde{V}_{de}
\end{array}
\right)_{\mbox{adiabatic}} = \textbfss{C}_3 \left(
\begin{array}{c}
3/4  \\
-(5/4) \mathcal{P} \\
1 \\
- (5/4) \mathcal{P}\\
3/4  \\
1 \\
- (5/4) \mathcal{P}\\
- \mathcal{P} \\
 1/4  \\
 - (5/4) \mathcal{P}
\end{array}
\right) \,,
\end{equation}
where $\mathcal{P} = \left( 15 + 4\,R_\nu \right)^{-1}$. This
corresponds to equations (59--61) of \citet{Valiviita:2008iv}. All
components except $\Delta_{de}$ are equal to the initial
conditions for $-2/3 < w_e < 1/3$, Eq.~(\ref{eq.adiabatic_mode}).
However, as pointed out in \citet{Valiviita:2008iv}, $\Delta_{de} =
\Delta_\gamma /4$ corresponds exactly to the adiabaticity
condition for DE: $S_{de\,A}=0$. Namely, substituting the result
(\ref{eq:departinS}) into definition (\ref{eq.entropy}), we find
\be S_{de\,\gamma} = 3 \Delta_{de} - \frac{3}{4} \Delta_\gamma \,.
\ee Thus Eq. (\ref{eq.adiabatic_mode_w<-2/3}) is an adiabatic
initial condition vector for the cases $-4/5 \le w_e \le -2/3$ or
$w_e < -1$.

\section{Conclusion}

We have presented, for the first time, a \emph{systematic}
derivation of initial conditions for perturbations in a model of
interacting dark matter - dark energy fluids, in the early
radiation era.
These initial conditions are essential for studying
the further evolution of perturbations up to today's observables.
They are the initial values for perturbations in any Boltzmann
integrator which solves the multipole hierarchy and produces the
theoretical predictions for the CMB temperature and polarization
angular power spectrum, as well as the matter power spectrum. We have
focused on the interaction $Q_c^\mu =- \Gamma\rho_c (1+\delta_c)
u_c^\mu $, where $\Gamma$ is a constant rate of energy density
transfer [see Eqs.~(\ref{eq.cont_rhoc}) and
(\ref{eq.cont_rhode})]. Generalising a previous result for
non-interacting dark energy in \citet{Doran:2003xq}, we find that,
in our interacting model, requiring adiabaticity between all the
other constituents (photons, neutrinos, baryons, and cold dark
matter) leads automatically also to dark energy adiabaticity, if
its early-time equation of state parameter is $w_e < -1$ or $-4/5
\le w_e \le 1/3$. In our previous work \citep{Valiviita:2008iv}, we
showed that if the equation of state parameter for dark energy is
$-1 < w_{de} < -4/5$ in the radiation or matter eras, the model
suffers from a serious non-adiabatic instability on large scales.
In this paper, the systematic derivation of initial conditions
confirms that result. However, in this paper we have shown that
the instability can easily be avoided, if we allow for suitably
time-varying dark energy equation of state.  The main results
are verbally summarised in Table~\ref{table:summary}
on page \pageref{table:summary}.

In the companion paper \citep{CompanionLike} we modified the
CAMB Boltzmann integrator\footnote{http://camb.info} \citep{Lewis:1999bs},
using the adiabatic initial conditions derived here for the
interacting model, and performed full Monte Carlo Markov Chain 
likelihood scans for this model as well as for the non-interacting
($\Gamma$=0) model for a reference, with various combinations of
publicly available data sets: WMAP \citep{Komatsu:2008hk}, WMAP \& ACBAR \citep{Reichardt:2008ay}, SN \citep{Kowalski:2008ez}, BAO \citep{Percival:2007yw},
WMAP\&SN, WMAP\&BAO, WMAP\&SN\&BAO.

With the parametrization $w_{de} = w_0a + w_e (1-a)$,
viable interacting cosmologies result for
$w_0$ close to $-1$ and $w_e < -1$ or $-4/5 < w_e \le 1/3$, as
long as $w_0+1$ and $w_e+1$ have the same sign \citep{CompanionLike}.
These particular conclusions apply exclusively to the  interaction model
we considered in this paper.

However, the method can be easily adapted
for studying different interactions: one only needs to modify the background evolution and interaction terms in Eqs.~(\ref{eq.Dc_newnew}), (\ref{eq.Vc_newnew}),
(\ref{eq.Dx_newnew}), and (\ref{eq.Vx_newnew}), before reading a new
matrix $\textbfss{A}(x)$ from them.
Based on section IV of \citet{Valiviita:2008iv}, the other interacting
fluid models
[$aQ_c=-\alpha \h \rho_c$ or $aQ_c=-\beta \h (\rho_c+\rho_{de})$, where
$\alpha,\;\beta \lesssim 1$ are dimensionless constants],
that are common in the literature, behave in a very similar
way to the model studied here, i.e., for $-1 < w_e < w_{\rm crit}$ the
models are not viable due to the early-time large-scale blow-up of
perturbations, for $w_{\rm crit} < w_e < w_{\rm adiab}$ the models
can be viable and \emph{non-standard} adiabatic initial conditions maybe found,
and for $w_e >  w_{\rm adiab}$ (or $w_e<-1$) the models are viable and 
standard (non-interacting) adiabatic initial conditions can be found.
The critical value $w_{\rm crit}$ is determined by demanding that
the 'blow-up' mode is actually a decaying mode and the fastest 'growing'
curvature perturbation mode is a constant, i.e., that the largest real part
of the eigenvalues $\lambda_i$ is zero, which with the notation
of \citet{Valiviita:2008iv} is guaranteed whenever ${\rm Re}(n_+) \leq 3$.
In our model the critical value, $w_{\rm crit} = -4/5$, is independent of the
strength of interaction, but in the above-mentioned models
it depends on $\alpha$ or $\beta$, as indicated by equations (85) and
(98) in \citet{Valiviita:2008iv}.
In general, our results show that the (early-time) dark energy equation of state plays, 
together with the interaction model, an important role in the (in)stability
of perturbations.

\vspace{2mm} {\bf Acknowledgments:} JV and RM are supported by
STFC. During this work JV received support also from the Academy
of Finland.

\bibliography{InteractingDEperturbations}
\bibliographystyle{mn2e}

\label{lastpage}
\end{document}